\documentclass[aps,prd,nofootinbib,twocolumn,floatfix,superscriptaddress,secnumarabic]{revtex4}
\pdfoutput=1
\usepackage{amsfonts,amssymb,epsfig,verbatim,mathbbol,amstext,amsmath,psfrag}
\usepackage{graphicx}
\usepackage[bookmarks,
                 bookmarksopen = true,
                 bookmarksnumbered = true,
                 linktocpage,
                 colorlinks = true,
                 linkcolor = blue,
                 urlcolor  = blue,
                 citecolor = blue,
                 anchorcolor = green,
                 hyperindex = true,
                 hyperfigures]
		{hyperref}
\usepackage[latin2]{inputenc}
\usepackage{t1enc}
\usepackage{amsmath}
\usepackage{subfigure}
\usepackage{dsfont}
\usepackage{slashed}
\usepackage{multirow}
\usepackage{lscape}
\usepackage{wrapfig}
\usepackage{array}

\usepackage{setspace}

\usepackage{colordvi}
\usepackage{lscape}
\usepackage{rotfloat}
\usepackage{rotating}
\usepackage{color}

\usepackage{cmap}

\newcommand{\bdm}{\begin{dmath}}
\newcommand{\edm}{\end{dmath}}
\newcommand{\bdms}{\begin{dmath*}}
\newcommand{\edms}{\end{dmath*}}
\newcommand{\bdg}{\begin{dgroup*}}
\newcommand{\edg}{\end{dgroup*}}

\def\lsim{\mathrel{\rlap{\lower4pt\hbox{\hskip1pt$\sim$}}
    \raise1pt\hbox{$<$}}}                
\def\gsim{\mathrel{\rlap{\lower4pt\hbox{\hskip1pt$\sim$}}
    \raise1pt\hbox{$>$}}}                

\def\smn{{\sigma_{\mu\nu}}}

\def\openone{\mathds{1}}

\def\Zq{Z_{\rm q}}

\def\ZS{Z_{\rm S}}

\def\ZT{Z_{\rm T}}

\newcommand{\bea}{\begin{eqnarray}} 
\newcommand{\eea}{\end{eqnarray}}

\newcommand{\D}{\mathcal{D}} 

\newcommand{\be}{\begin{equation}}
\newcommand{\ee}{\end{equation}}

\newcommand{\tr}{\textmd{tr}}

\newcommand{\ZZ}{\mathcal{Z}}
\newcommand{\expv}[1]{\left \langle #1 \right \rangle}
\newcommand{\TAU}{\tau_f}

\long\def\symbolfootnote[#1]#2{\begingroup%
\def\thefootnote{\fnsymbol{footnote}}\footnote[#1]{#2}\endgroup} 

\def\sign{\textmd{sign}}

\def\chiF{\chi}
\def\chiM{\xi}
\def\dsf{\slashed{D}_f}
\def\MSBar{\overline{\rm MS}}
\newcommand{\Grnf}{Green }

\begin{document}

\title{Magnetic susceptibility of QCD at zero and at finite temperature from the lattice}

\author{G.~S.~Bali}
\affiliation{Institute for Theoretical Physics, Universit\"at Regensburg, D-93040 Regensburg, Germany}
\affiliation{Department of Theoretical Physics, Tata Institute of Fundamental Research, Homi Bhabha Road, Mumbai 400005, India.}
\author{F.~Bruckmann}
\affiliation{Institute for Theoretical Physics, Universit\"at Regensburg, D-93040 Regensburg, Germany}
\author{M.~Constantinou}
\affiliation{Department of Physics, University of Cyprus, Nicosia, CY-1678, Cyprus}
\author{M.~Costa}
\affiliation{Department of Physics, University of Cyprus, Nicosia, CY-1678, Cyprus}
\author{G.~Endr\H{o}di}
\thanks{\,Corresponding author. \href{mailto:gergely.endrodi@physik.uni-r.de}{gergely.endrodi@physik.uni-r.de}}
\affiliation{Institute for Theoretical Physics, Universit\"at Regensburg, D-93040 Regensburg, Germany}
\author{S.~D.~Katz}
\affiliation{Institute for Theoretical Physics, E\"otv\"os University, H-1117 Budapest, Hungary}
\author{H.~Panagopoulos}
\affiliation{Department of Physics, University of Cyprus, Nicosia, CY-1678, Cyprus}
\author{A.~Sch\"afer}
\affiliation{Institute for Theoretical Physics, Universit\"at Regensburg, D-93040 Regensburg, Germany}

\begin{abstract}
The response of the QCD vacuum to a constant external (electro)magnetic field is studied through the tensor polarization of the chiral condensate and the magnetic susceptibility at zero and at finite temperature. We determine these quantities using lattice configurations generated with the tree-level Symanzik improved gauge action and $N_f=1+1+1$ flavors of stout smeared staggered quarks with physical masses. We carry out the renormalization of the observables under study and perform the continuum limit both at $T>0$ and at $T=0$, using different lattice spacings. Finite size effects are studied by using various spatial lattice volumes. The magnetic susceptibilities $\chi_f$ reveal a spin-diamagnetic behavior; we obtain at zero temperature $\chi_u=-(2.08\pm0.08) \textmd{ GeV}^{-2}$, $\chi_d=-(2.02\pm0.09) \textmd{ GeV}^{-2}$ and $\chi_s=-(3.4\pm1.4) \textmd{ GeV}^{-2}$ for the up, down and strange quarks, respectively, in the $\MSBar$ scheme at a renormalization scale of $2 \textmd{ GeV}$.
We also find the polarization to change smoothly with the temperature in the confinement phase and then to drastically reduce around the transition region.

\end{abstract}


\maketitle

\section{Introduction}
\label{sec:intro}

An external (electro)magnetic field is an excellent probe of the dynamics of the QCD vacuum. Strong magnetic fields affect fundamental properties of QCD like chiral symmetry breaking and restoration, deconfinement, the hadron spectrum or the phase diagram, just to name a few. 
Chiral symmetry breaking has long been known to be enhanced by magnetic fields at zero temperature, signalled by an increasing chiral condensate (see e.g. Ref.~\cite{Gusynin:1995nb}). The particle spectrum may undergo drastic changes (see e.g. the ongoing discussion in Refs.~\cite{Chernodub:2010qx,Hidaka:2012mz,Chernodub:2012zx}) with some strong decay channels becoming unavailable and others opening up. 
The transitions at non-vanishing temperature related to chiral symmetry breaking and deconfinement are also affected by the magnetic field $B$. The phase diagram of QCD in the temperature-magnetic field plane was determined recently in lattice simulations~\cite{Bali:2011qj,Bali:2011uf,Bali:2012zg} by analyzing the dependence of the chiral condensate and of other observables on $B$, with the main result that the transition temperature $T_c$ decreases with growing $B$\footnote{Employing physical quark masses in the simulation and extrapolating the results to the continuum limit, as was done in Refs.~\cite{Bali:2011qj,Bali:2011uf,Bali:2012zg}, proved to be essential. Studies where these ingredients are missing produce qualitatively different results, namely an increasing $T_c(B)$ function~\cite{D'Elia:2010nq,Ilgenfritz:2012fw}. A possible explanation for this discrepancy and a comparison to effective theories was given recently in Ref.~\cite{Bali:2012zg}.
} and the transition remains an analytic crossover just as at $B=0$~\cite{Aoki:2006we}.
These effects are relevant in several physical situations as strong magnetic fields are expected to play a significant role, e.g., in early cosmology~\cite{Vachaspati:1991nm}, in non-central heavy ion collisions~\cite{Skokov:2009qp} and in dense neutron stars~\cite{Duncan:1992hi}.

Another fundamental characteristic of the QCD vacuum is the response of the free energy density (which at zero temperature is the vacuum energy density) to magnetic fields,
\be
f=-\frac{T}{V} \log \ZZ,
\label{eq:freeenergy}
\ee
where $\ZZ$ is the partition function of the system and $V$ the (three-dimensional) volume. Due to rotational invariance the $B$-dependence of $f$ is to leading order quadratic, characterized by the magnetic susceptibility of the QCD vacuum,
\be
\chiM = -\left.\frac{\partial^2 f}{\partial (eB)^2}\right|_{eB=0},
\label{eq:defchitotal}
\ee
which is a dimensionless quantity (here $e>0$ denotes the elementary charge).
A positive susceptibility indicates a decrease in $f$ due to the magnetic field, that is to say, a {\it paramagnetic} response. On the other hand $\chiM<0$ is referred to as \emph{diamagnetism}~\cite{springerlink:10.1007/BF01397213}.
Clearly, the sign of $\chiM$ is a fundamental property of the QCD vacuum.

In the functional integral formalism of QCD the susceptibility is readily split into spin- and orbital angular momentum-related terms, according to
\be
\chiM=\sum_f\chiM_f,\quad\quad \chiM_f=\chiM^S_f+\chiM^L_f,
\label{eq:chiSO}
\ee
with contributions from each quark flavor $f$ with electric charge $q_f$ and mass $m_f$. For a constant magnetic field $B=F_{xy}$ in the positive $z$ direction,
\be
 \chiM^S_f = \frac{q_f/e}{2m_f}  
 \left.\frac{\partial} {\partial (eB)}\expv{\bar\psi_f\sigma_{xy}\psi_f}\right|_{eB=0},
 \quad\smn = \frac{1}{2i} [\gamma_\mu,\gamma_\nu ].
\label{eq:therelation}
\ee
$\chiM^L_f$ is given by an analogous expression with $\sigma_{xy}$ replaced by 
a generalized angular momentum also present for spinless particles, cf.\ Eq.~(\ref{eq:xicontr}) of Appendix~\ref{app:logZB}. 
Eq.~(\ref{eq:therelation}) constitutes an important relation which, to our knowledge, has not been recognized previously in this context. Its derivation from the quark determinant and the corresponding Dirac operator is given in Appendix~\ref{app:logZB}.

In the present paper we concentrate on the spin contributions, and thus the expectation value of the tensor polarization operator $\bar\psi_f \smn \psi_f$. To leading order this is proportional to the field strength and thus can be written as~\cite{Ioffe:1983ju}
\be
\expv{ \bar\psi_f \sigma_{xy} \psi_f } = q_f B \cdot \expv{\bar \psi_f\psi_f}\cdot \chiF_f \equiv q_f B \cdot \TAU,
\label{eq:defchi}
\ee
where $\expv{\bar \psi_f\psi_f}$ is the expectation value of the quark condensate (see Eq.~(\ref{eq:pbpdef}) below for its definition). Corrections to the right hand side are expected to be of $\mathcal{O}(B^3)$, so that Lorentz invariance is maintained. In the literature $\chiF_f$ is referred to as the \emph{magnetic susceptibility of the condensate} (for the quark flavor $f$). In what follows we will also use the term ``magnetic susceptibility''. Again we stress that it constitutes only one of the two contributions to the total susceptibility. We also define the \emph{tensor coefficient} $\TAU$ as the product of the condensate and the magnetic susceptibility. Both quantities will depend on the temperature $T$ at which the expectation values of Eq.~(\ref{eq:defchi}) are determined. At finite quark masses it is advantageous to work with $\TAU$ instead of $\chiF_f$ for reasons related to renormalization (see below). 

The magnetic susceptibility $\chiF_f$, in the context of QCD, was first introduced in Ref.~\cite{Ioffe:1983ju}. 
Since then its experimental relevance has been growing steadily.
In particular, this quantity appears in the description of radiative $D_s$ meson transitions~\cite{Colangelo:2005hv}, of the anomalous magnetic moment of the muon~\cite{Czarnecki:2002nt} and of chiral-odd photon distribution amplitudes~\cite{Braun:2002en,Pire:2009nn}.
Moreover, vector-tensor two-point functions at zero momentum are related to the magnetic susceptibility~\cite{Nyffeler:2009uw}. 

Since $\chiF_f$ acts as an input parameter in various strong interaction processes~\cite{Rohrwild:2007yt}, 
a high-precision determination of its value is of importance. In the past, the magnetic susceptibility has been calculated using QCD sum rules~\cite{Belyaev:1984ic,Balitsky:1985aq,Ball:2002ps}, in the holographic approach~\cite{Bergman:2008sg,Gorsky:2009ma}, using the operator product expansion~\cite{Vainshtein:2002nv}, in the instanton liquid model and chiral effective models~\cite{Kim:2004hd,Dorokhov:2005pg,Goeke:2007nc,Nam:2008ff}, using the zero modes of the Dirac operator~\cite{Ioffe:2009yi}, and in low-energy models of QCD like the quark-meson model and the Nambu-Jona-Lasinio (NJL) model~\cite{Frasca:2011zn}. The numerical value of $\chiF_f$ was also determined recently on the lattice in the quenched approximation of two-~\cite{Buividovich:2009ih} and of three-color QCD~\cite{Braguta:2010ej}, in both cases without renormalization. We mention that the quenched approach can lead to large systematic errors at strong magnetic fields~\cite{D'Elia:2011zu}.

In this paper we determine $\TAU(T)$ and $\chiF_f(T)$ for a wide range of temperatures around the transition region between the hadronic and the quark-gluon plasma phases and at $T=0$. We apply fully dynamical lattice simulations, i.e. both the fermionic degrees of freedom and the external field are taken into account in the generation of the gauge ensembles. We perform the renormalization of the tensor coefficient and carry out the continuum extrapolation using results obtained at different lattice spacings.
One main result will be that the tensor coefficients at $T=0$ are negative, indicating the spin-diamagnetic nature of the QCD vacuum. Moreover we observe that $\tau_f$ decreases around the QCD crossover temperature similarly to other order parameters like the condensate.

This paper is organized as follows.
We define the lattice implementation of the magnetic field and the observables in Sec.~\ref{sec:magnfield} and discuss their renormalization in Sec.~\ref{sec:renormalization}. The multiplicative renormalization is carried out perturbatively; the determination of renormalization constants is detailed in Sec.~\ref{sec:operators}. After a brief summary of the simulation setup in Sec.~\ref{sec:simdet} we present the results in Sec.~\ref{sec:results} for the tensor coefficients and in Sec.~\ref{sec:magnsusc} for the susceptibilities, before we conclude.

\section{Magnetic field and observables}
\label{sec:magnfield}

We study the effect of an external magnetic field $B$ on the expectation value of the tensor polarization, Eq.~(\ref{eq:defchi}). To realize such an external field on the lattice we implement the continuum $\mathrm{U}(1)$ gauge field $A_\mu$ satisfying $\partial_x A_y - \partial_y A_x = B$ using space-dependent complex phases~\cite{Martinelli:1982cb,Bernard:1982yu,D'Elia:2010nq,Bali:2011qj} in the following way,
\be
\begin{split}
u_y(n) &= e^{i a^2 q_f B n_x}, \\
u_x(N_x-1,n_y,n_z,n_t) &= e^{-i a^2 q_f B N_x n_y}, \\
u_x(n) &= 1, \quad\quad\quad\quad n_x\ne N_x-1, \\
u_\nu(n) &=1, \quad\quad\quad\quad \nu\not\in\{x,y\},
\end{split}
\label{eq:links2}
\ee
where the sites are labeled by integers $n\equiv(n_x,n_y,n_z,n_t)$, with $n_\nu=0\ldots N_\nu-1$ and $a$ is the lattice spacing.
This prescription for the links corresponds to a covariant derivative for the flavor $f$ of the form $D_{\mu,f}=\partial_\mu+iq_fA_\mu + iA_\mu^a T^a$
\footnote{Note that we do not include in the action the corresponding photon kinetic term $F_{\mu\nu}F_{\mu\nu}/4=B^2/2$. This means that in the discussion we will never encounter $B$ alone but only the combination $q_fB\sim eB$.}.
This discretization satisfies periodic boundary conditions in the spatial directions and ensures that the magnetic flux across the $x-y$ plane is constant. It is well known that the magnetic flux in a finite volume is quantized~\cite{'tHooft:1979uj,AlHashimi:2008hr}, which on the lattice implies
\be
qB\cdot a^2 = \frac{2\pi N_b}{N_x N_y}, \quad N_b\in\mathds{Z},\quad 0\le N_b < N_xN_y,
\label{eq:quantlatt}
\ee
where $q$ is the smallest charge in the system, in our case $q=q_d=q_s=-e/3$. Due to the periodicity of the links of Eq.~(\ref{eq:links2}) in $N_b$ with period $N_xN_y$, one expects lattice artefacts to become large if $N_b>N_xN_y/4$. In the following we use lattices with $N_x=N_y=N_z\equiv N_s$.

We consider three quark flavors $u,d$ and $s$. Since the charges and masses of the quarks differ we have to treat each flavor separately; $q_u=-2q_d=-2q_s$. We assume $m_u=m_d\ne m_s$. The partition function in the staggered formulation then reads,
\be
\ZZ = \int \D U e^{-\beta S_g} \prod_{f=u,d,s} \left[ \det M(U,q_fB,m_f)\right]^{1/4}, 
\label{eq:partfunc}
\ee
with $M(U,qB,m) = \slashed D(U,qB) + m\mathds{1}$ being the fermion matrix and $\beta=6/g^2$ the gauge coupling. The exact form of the action we use is described in Refs.~\cite{Aoki:2005vt,Borsanyi:2010cj}, and further details of the simulation setup are given in Sec.~\ref{sec:simdet}. Since the external field couples directly only to quarks, $B$ just enters the fermion determinants, through the $\mathrm{U}(1)$ links of Eq.~(\ref{eq:links2}). The volume of the system is given as $V\equiv(aN_s)^3$ and the temperature as $T=(aN_t)^{-1}$.

In this formulation the expectation value of the quark condensate for the flavor $f$ can be written as
\be
\expv{\bar{\psi}_f\psi_f} \equiv \frac{T}{V}\frac{\partial \log \ZZ}{\partial m_f} = \frac{T}{4V} \expv{\tr\, M^{-1}(U,q_fB,m_f)}.
\label{eq:pbpdef}
\ee
Likewise, the expectation value of the tensor Dirac structure reads,
\be
\expv{\bar\psi_f \smn \psi_f} = \frac{T}{4V} \expv{\tr \, M^{-1}(U,q_fB,m_f)\smn}.
\label{eq:pbpTdef}
\ee
The staggered realization of this operator is detailed in Appendix~\ref{app:staggeredops}.

At this point a few comments regarding the sign of the expectation values in Eq.~(\ref{eq:defchi}) are in place. In continuum calculations a negative sign for the condensate is customary, see e.g. Ref.~\cite{Ball:2002ps}, in contrast to our convention in Eq.~(\ref{eq:pbpdef}). This sign convention applies for any fermionic bilinear expectation value, therefore it does not affect the sign of $\chiF_f$, but only that of $\TAU$. Further possible differences in the sign can arise from the definition of $\smn$ and from that of the $\mathrm{U}(1)$ part of the covariant derivative.
We note that our notation is consistent with that of Ref.~\cite{Ball:2002ps} in terms of $\smn$, but differs by a minus sign in the covariant derivative (see the paragraph below Eq.~(\ref{eq:links2})), implying an overall relative minus sign of $\chiF_f$.

\section{Renormalization}
\label{sec:renormalization}

In order to determine the continuum limit of the observables defined in Eqs.~(\ref{eq:pbpdef}) and~(\ref{eq:pbpTdef}), their renormalization has to be performed.
The quark condensate (at finite mass) is subject to additive and multiplicative renormalization, due to the divergent terms in the free energy density $f$ of Eq.~(\ref{eq:freeenergy}) and in the bare mass $m_f$. The former divergence is (to leading order) quadratic in the cutoff $1/a$~\cite{Leutwyler:1992yt}. 
Therefore, the bare observable can be written as
\be
\expv{\bar\psi_f\psi_f}(B,T) = \frac{1}{\ZS}\expv{\bar\psi_f\psi_f}^r(B,T) + \zeta_S m_f/a^2+\ldots,
\label{eq:renS}
\ee
where $\ZS$ is the renormalization constant of the scalar operator and the ellipses denote subleading (logarithmic) divergences in $a$. Here the superscript $r$ indicates the renormalized observable. The divergences in $\expv{\bar\psi\psi}$ depend neither on the temperature nor on the external field\footnote{For a detailed argumentation about the absence of $B$-dependent divergences in the condensate see Ref.~\cite{Bali:2011qj} and references therein.}. Therefore, in mass-independent renormalization schemes, $\zeta_S$ and $\ZS$ are just functions of the gauge coupling.
The conventional way to cancel the additive divergences is to consider the difference, for example, between the condensate at $T$ and at $T=0$.

The situation is somewhat different for the tensor polarization.
As a calculation in the free theory shows, an additive divergence of the form $q_fB m_f \log(m_f^2a^2)$ appears in $\expv{\bar\psi\smn\psi}$ (see Appendix~\ref{app:mlogm}). This divergence vanishes in the chiral limit (or at zero external field) and is not related to the multiplicative divergence of the tensor operator to which we will return below. Altogether the bare observable can thus be written as
\be
\begin{split}
\expv{\bar\psi_f\sigma_{\mu\nu}\psi_f}(B,T) & \\
&\hspace*{-2.6cm}=\frac{1}{\ZT} \expv{\bar\psi_f\sigma_{xy}\psi_f}^r(B,T) + \zeta_T q_fB m_f\log(m_f^2a^2)+\ldots,
\end{split}
\label{eq:renT}
\ee
where $\ZT$ is the renormalization constant of the tensor operator (its perturbative determination is detailed in Sec.~\ref{sec:operators}) and $\zeta_T$ the coefficient of the divergent logarithm. Both are independent of $T$ and $B$ (and in mass-independent schemes of $m_f$). In Eq.~({\ref{eq:renT}}) the ellipses denote finite terms.
In the free theory we calculate $\zeta_T(g=0)=3/(4\pi^2)$ (see Appendix~\ref{app:mlogm}).

 From these considerations it is clear that the magnetic susceptibility $\chiF_f$, being proportional to the ratio of Eq.~(\ref{eq:renT}) over Eq.~(\ref{eq:renS}), at non-vanishing quark mass contains additive divergences which depend both on $T$ and on $B$ (and also on the quark flavor $f$). This means that these singular contributions cannot be removed by subtracting the same operator, measured at different $T$ or $B$ (or flavor $f$).

Therefore, in the following we consider the tensor coefficient $\TAU$ defined in Eq.~(\ref{eq:defchi}). We notice that the operator $1-m_f\partial/\partial m_f$ eliminates the logarithmic divergence and thus can be used to define an observable with a finite continuum limit,
\be
\TAU^r \equiv \left(1-m_f\frac{\partial}{\partial m_f}\right)\TAU \cdot \ZT \equiv \TAU \ZT - \TAU^{\rm div}.
\label{eq:mdmsub}
\ee
At finite quark mass this is one possible prescription to cancel the additive logarithmic term. 
It has the advantages that the chiral limit of $\TAU$ is left unaffected, and that, together with the logarithmic divergence, scheme-dependent finite terms also cancel in this difference (see Eq.~(\ref{eq:polarfreefinal})), such that the scheme- and renormalization scale-dependence of $\TAU$ resides solely in $\ZT$.

Since the subtracted divergence is independent of the temperature, we are able to determine $\TAU^{\rm div}$ at zero temperature where we systematically study the dependence of $\TAU$ on $m_f$ and $a$, and then to perform the subtraction at nonzero temperatures as well. As we will see, the subtraction in Eq.~(\ref{eq:mdmsub}) amounts to a $5-10$ per cent effect for the lattice spacings we use. 

\section{Perturbative evaluation}
\label{sec:operators}


\subsection{Fermion propagator and fermion bilinears}

\begin{figure}[ht!]
\centering
\includegraphics*[height=2.0cm]{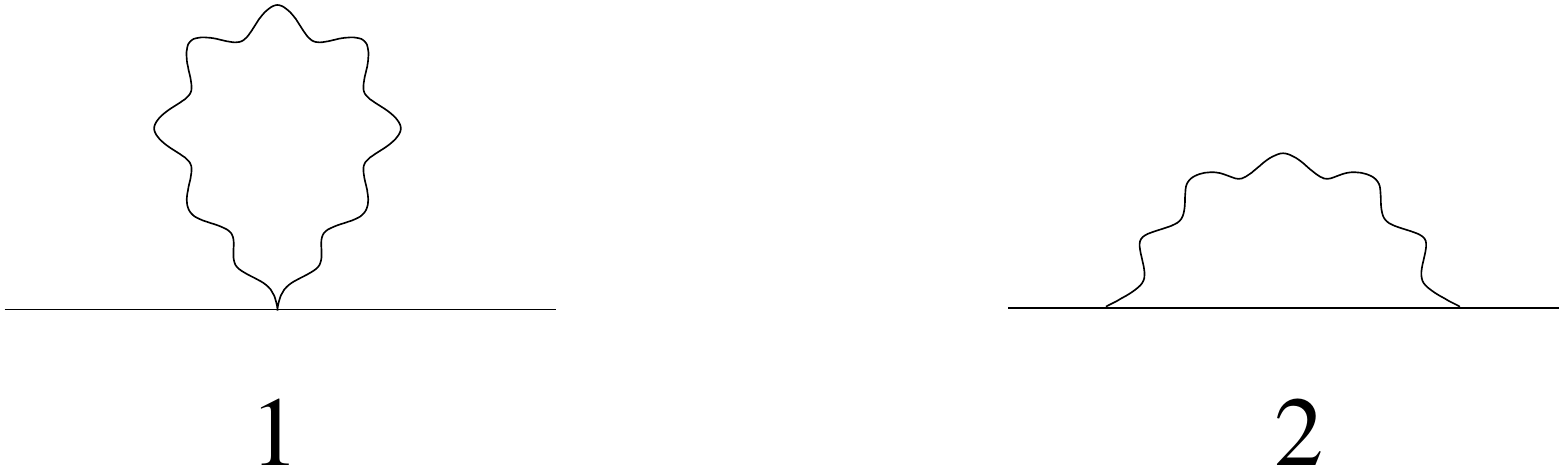}
\caption{One-loop diagrams contributing to the
fermion propagator. Wavy (solid) lines represent gluons (fermions).}
\label{figprop1}
\end{figure}

The discretization of the continuum bilinear operators in the staggered formulation of lattice QCD is detailed in Appendix~\ref{app:staggeredops}.
Using this discretization we compute the one-loop correction to the fermion propagator in order to
obtain its renormalization constant, an essential ingredient for the
renormalization of fermion operators defined in Eqs.~(\ref{eq:OS2}) and~(\ref{eq:OT2}).
The one-loop Feynman diagrams that enter our two-point amputated \Grnf function
calculation for the propagator are illustrated in Fig.~\ref{figprop1}. 
For the algebraic operations involved in evaluating Feynman
diagrams, we make use of our symbolic package in Mathematica.
The required procedure for the computation of a Feynman diagram
can be found in Ref.~\cite{Constantinou:2009tr}.

We provide the total expression for the inverse fermion
propagator $S^{-1}(p)$, computed up to one-loop in
perturbation theory. Here we should point out that, for
dimensional reasons, there is a global prefactor $1/a$ multiplying our
expressions for the inverse propagator, and thus, the ${\cal O}(a^0)$
correction is achieved by considering all terms up to ${\cal O}(a^1)$.
We have computed $S^{-1}(p)$ for general values of the gauge parameter $\alpha$,
the stout smearing parameters $\omega_1,\omega_2$ (since we perform 
two stout steps, we have kept these different, so that our results are 
also applicable for the single stout smearing case), the bare quark masses
and the external momentum. We have obtained results using 10 sets 
of values for the Symanzik improvement coefficients of the gluon action 
(shown in Ref.~\cite{Constantinou:2009tr}). In presenting our result for
$S^{-1}(p)$ below, we use the Landau gauge for conciseness; the quantities
$c_i$ are numerical coefficients that depend on the gluon 
action we choose and on the stout smearing parameters.

\begin{figure}[ht!]
\centering
\includegraphics*[height=2.8cm]{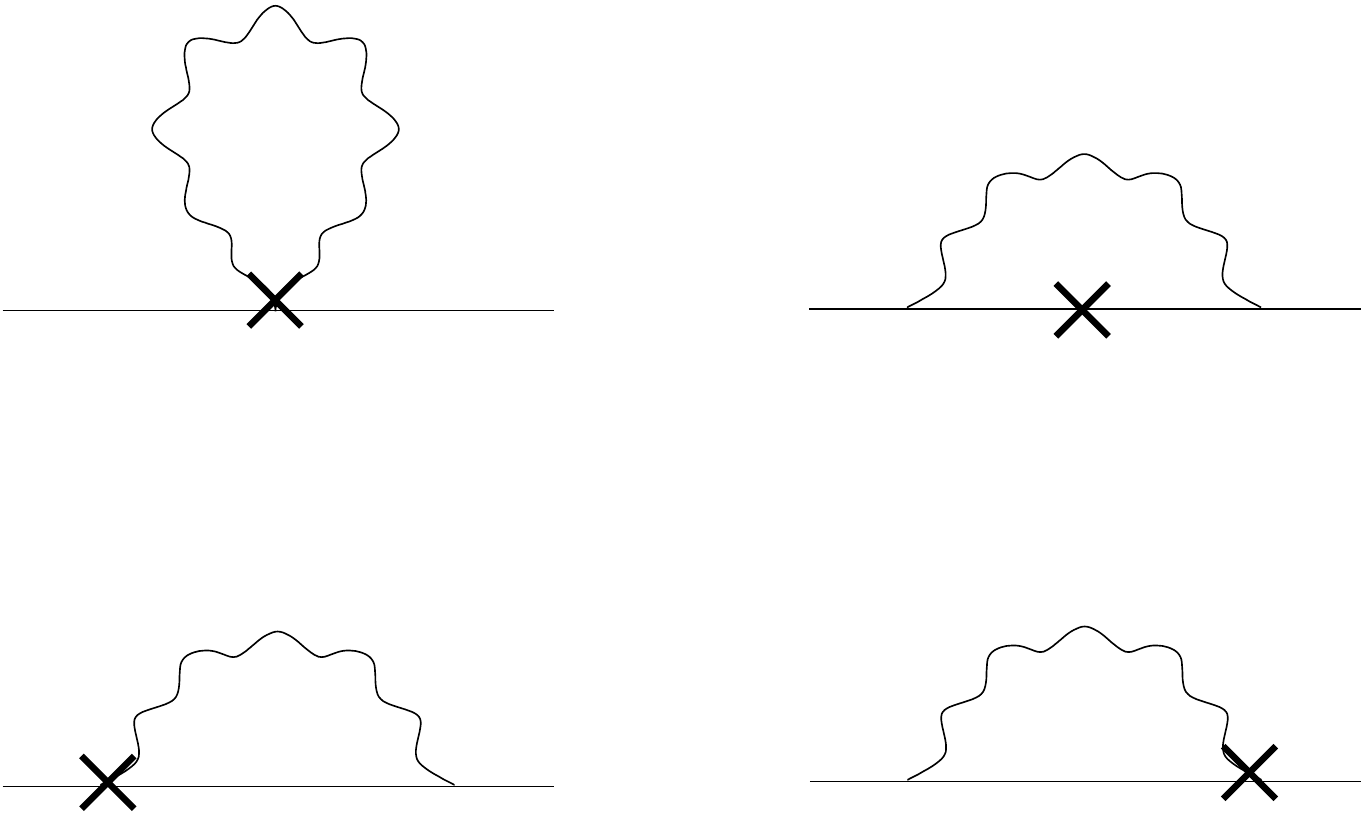}
\caption{One-loop diagrams contributing to the 
bilinear operators. A wavy (solid) line represents gluons (fermions). A
cross denotes the insertion of a Dirac matrix.}
\label{figbil2}
\end{figure}

\begin{widetext}
\vspace*{-0.5cm}
\bea
S^{-1}_{\rm 1-loop} &=&
\left(\sum_\rho\,\delta\left(q_1 - q_2 + \frac{\pi}{a}\bar{\rho}\right)\,i\,p_\rho\right)\left(
        1 + \frac{g^2\,C_F}{16 \pi^2}\,\left( c_1 + c_2 \log\left(a^2 m^2
        + a^2 p^2\right)+ c_3 \frac{m^2}{p^2}+ c_4
        \frac{m^4}{p^4}\,\log\left(1+\frac{p^2}{m^2}\right)
        \right)\right) \nonumber \\
&+&\delta(q_1 - q_2) m \left( 1 +
        \frac{g^2\,C_F}{16 \pi^2}\,\left( c_6 + c_7 \log\left(a^2 m^2 +
        a^2 p^2\right)+ c_8
        \frac{m^2}{p^2}\,\log\left(1+\frac{p^2}{m^2}\right)\right)
        \right),
\eea
\end{widetext}
\vspace*{-0.6cm}
where
\be
\nonumber
C_F \equiv \frac{N_c^2-1}{2N_c}\,,\quad\quad\quad \bar{\rho}=\sum_{\nu=1}^{\rho-1}\hat{\nu}\,,
\ee
\vspace*{-0.5cm}
\be
\nonumber
p=\left(q_1+\frac{\pi}{2}\right)_{{\rm mod}\,\pi} - \frac{\pi}{2} = \left(q_2+\frac{\pi}{2}\right)_{{\rm mod}\,\pi} - \frac{\pi}{2}\,,
\ee
$N_c=3$ is the number of colors and $q_1$ and $q_2$ are the external momenta.
Next we study the one-loop perturbative
\Grnf functions of the tensor fermionic operator.
The one-particle irreducible Feynman diagrams that enter the
calculation of the above \Grnf functions are shown in
Fig.~\ref{figbil2}, and include up to two-gluon vertices extracted from
the operator (the cross in the diagrams). The appearance of gluon
lines attached to the operator stems from the product of gauge links in the
operator definition (see Eq.~(\ref{eq:Oper})).


\subsection{Quark field and quark bilinear renormalization constants in
the RI$'$-MOM scheme}

The renormalization constant (RC) can be thought of as the link
between the matrix element, regularized on the lattice (without upper index), and the
renormalized continuum counterpart (with upper index $r$). For the fermion field, the scalar and the tensor operator this is written as
\be
\psi^r = \Zq^{1/2}\, \psi\,,\quad 
{\cal O}^r_S = \ZS\, {\cal O}_S\,,\quad
{\cal O}^r_T = \ZT\,  {\cal O}_T\,.
\ee
The RCs of lattice operators are necessary ingredients in the
prediction of physical probability amplitudes from lattice matrix
elements. In this section we present the multiplicative RCs, in the
RI$'$-MOM scheme, of the quark field ($\Zq$) and the quark bilinear operators ($\ZS$ and $\ZT$).

In the RI$'$-MOM renormalization scheme the forward
amputated \Grnf function $\Lambda(p)$, computed in the chiral
limit and at a given (large Euclidean) scale $p^2=\mu^2$, is imposed to be equal to
its tree-level value. The RCs are computed at arbitrary values of the
renormalization scale. 
\be
\begin{split}
S_{\rm 1-loop}^{-1}(p=\mu,m=0)  &= S_{\rm tree}^{-1}(\mu,m=0) \Zq(\mu), \\
\Lambda^{s}_{\rm 1-loop}(p=\mu,m=0) &= \Lambda^{s}_{\rm tree}(\mu,m=0) \Zq(\mu)\,Z_s^{-1}(\mu),
\end{split}
\nonumber
\ee
where $s=S$ corresponds to the scalar and $s=T$ to the tensor bilinear operator. Moreover, here $\mu$ is the renormalization scale,
$S^{-1}_{\rm tree}$ is the tree-level result for the inverse propagator,
and $\Lambda^{s}_{\rm tree}$ is the tree-level value of the \Grnf function
for the operator under study.

The expressions we obtain using our one-loop results discussed in the
previous subsection are shown here only for the tree-level improved Symanzik gauge
action. We allow for a general gauge characterized by the parameter $\alpha$
(Landau gauge: $\alpha=0$, Feynman gauge: $\alpha=1$) and stout smearing
parameters. As discussed above, we have employed different
parameters for the two smearing steps; in fact, we have also kept the
parameters of the action's smearing procedure ($\omega_{A_{1}},\omega_{A_{2}}$)
distinct from the parameters of the operator smearing 
($\omega_{O_{1}},\omega_{O_{2}}$). In all expressions the systematic
error (coming from an extrapolation to infinite lattice size of our
numerical loop-integrals) is smaller than the last digit we present. 
The results read:

\begin{widetext}
\vspace*{-0.6cm}
\bea
\hspace*{-.0cm}\Zq^{\rm RI'} = 1 + \frac{g^2C_F}{16 \pi^2}\Bigl[
&-&7.2136 + 4.7920\,\alpha + 124.5149\left({\omega_{A_{1}}}\! + {\omega_{A_{2}}}\right)
- 518.4332\left({\omega^2_{A_{1}}}\! + {\omega^2_{A_{2}}} \right)
- 2073.7329\,{\omega_{A_{1}}}{\omega_{A_{2}}} \nonumber \\ 
&+& 9435.3459\left({\omega^2_{A_{1}}}{\omega_{A_{2}}}
\!+ {\omega_{A_{1}}}{\omega^2_{A_{2}}}\right)
- 45903.1373\,{\omega^2_{A_{1}}}{\omega^2_{A_{2}}}
-\alpha \log\left(a^2\mu^2\right) \Bigr],
\eea

\vspace*{-0.6cm}
\bea
\hspace*{-.25cm}\ZS^{\rm RI'} = 1 + \frac{g^2C_F}{16 \pi^2}
\Bigl[&-&
34.3217 -\alpha + 389.2102 \left({\omega_{A_{1}}} + {\omega_{A_{2}}}\right) 
- 1403.6482\left({\omega_{A_{1}}}^2 + {\omega_{A_{2}}}^2 \right) 
- 5614.5930\,{\omega_{A_{1}}}{\omega_{A_{2}}}\nonumber\\
&+& 23395.3566\left({\omega_{A_{1}}}^2{\omega_{A_{2}}} + {\omega_{A_{1}}}{\omega_{A_{2}}}^2\right)
- 106813.9602 {\omega_{A_{1}}}^2{\omega_{A_{2}}}^2 
+ 3\log\left(a^2\mu^2\right) \Bigr] \label{eq:ZSRI}
\eea

\vspace*{-0.6cm}
\bea
\hspace*{-.05cm}\ZT^{\rm RI'} = 1 + \frac{g^2C_F}{16 \pi^2}\Bigl[&+&
8.8834 + \,\alpha\, + 116.5787\,\left({\omega_{A_{1}}} + {\omega_{A_{2}}}\right)
- 200.5879\,\left({\omega_{O_{1}}} + {\omega_{O_{2}}}\right)
- 531.7591\,\left({\omega^2_{A_{1}}} + {\omega^2_{A_{2}}} \right) \nonumber \\
&+& 780.5904\,\left({\omega^2_{O_{1}}} + {\omega^2_{O_{2}}} \right)
- 2095.1622\,{\omega_{A_{1}}}\,{\omega_{A_{2}}}
+ 3154.2357\,{\omega_{O_{1}}}\,{\omega_{O_{2}}} \nonumber \\
&+& 31.8743\,\left({\omega_{A_{1}}} + {\omega_{A_{2}}}\right)\,\left({\omega_{O_{1}}} + {\omega_{O_{2}}}\right)
+ 9877.2330\,\left({\omega^2_{A_{1}}}\,{\omega_{A_{2}}} + {\omega_{A_{1}}}\,{\omega^2_{A_{2}}}\right) \nonumber \\
&-& 13993.1045\,\left({\omega^2_{O_{1}}}\,{\omega_{O_{2}}} + {\omega_{O_{1}}}\,{\omega^2_{O_{2}}}\right)
- 284.0013\,\left({\omega_{A_{1}}} + {\omega_{A_{2}}}\right){\omega_{O_{1}}}\,{\omega_{O_{2}}} \label{eq:ZTRI} \\
&-& 284.0013\,{\omega_{A_{1}}}\,{\omega_{A_{2}}}\,\left({\omega_{O_{1}}} + {\omega_{O_{2}}}\right)
- 48519.2862\,{\omega^2_{A_{1}}}\,{\omega^2_{A_{2}}} \nonumber \\
&+& 68237.1178 \,{\omega^2_{O_{1}}}\,{\omega^2_{O_{2}}}     
+ 2709.4942 \,{\omega_{A_{1}}}\,{\omega_{A_{2}}}{\omega_{O_{1}}}\,{\omega_{O_{2}}} 
-\,\log\left(a^2\,\mu^2\right) \Bigr]. \nonumber
\hspace*{1.25cm}
\eea

\end{widetext}
\vspace*{-1.7cm}


\subsection{Conversion to the $\MSBar$ scheme}

Here we provide the expressions for the RCs in the
$\MSBar$ continuum scheme, using conversion factors
adapted from Refs.~\cite{Gracey:2003yr,Alexandrou:2012mt}. These conversion factors do
not depend on the regularization scheme (and, thus, they 
are independent of the lattice discretization), when expanded in terms of
the renormalized coupling constant. However, expressing them in terms
of the bare coupling constant in general introduces a dependence on the action.
To one-loop order, the renormalized and the bare couplings are nevertheless equal, leading to
\be
\Zq^{\MSBar} = \Zq^{\rm RI'} - \frac{g^2C_F}{16\pi^2} \alpha,
\ee
\vspace*{-.5cm}
\be
\ZS^{\MSBar} = \ZS^{\rm RI'} + \frac{g^2C_F}{16\pi^2} (4+\alpha),\quad
\ZT^{\MSBar} = \ZT^{\rm RI'} - \frac{g^2C_F}{16\pi^2} \alpha.
\label{eq:ZTMS}
\ee
We note that the RCs from the lattice scheme to the $\MSBar$ scheme are independent of the gauge parameter $\alpha$, as they should be.

The above conversion factors refer to the NDR (Naive Dimensional Regularization) version of $\MSBar$ (see, e.g., Ref.~\cite{Buras:1989xd}). Other possible modified minimal subtractions are related to NDR via additional finite renormalizations. In particular, to compare our result for $Z_T$ with the one in Ref.~\cite{Patel:1992vu}, which is given in the "DREZ" scheme (for the Wilson gauge action and without stout smearing), we must divide our NDR result by a factor $(1 + C_F g^2/(16\pi^2))$. The two results are in perfect agreement.

\section{Simulation setup}
\label{sec:simdet}

For our measurements we used the gauge ensembles of Refs.~\cite{Bali:2011qj,Bali:2012zg} augmented by additional new ensembles. All configurations were generated with the tree-level improved Symanzik gauge action and stout smeared staggered fermions, at physical quark masses. 
We use lattices at both $T=0$ and at $T>0$, at various values of the external magnetic field.
We employ two steps of stout smearing with parameter $\omega_{A}=\omega_{O}=0.15$ both in the action and in the operators. The zero temperature ensembles consist of $24^3\times 32$, $32^3\times 48$ and $40^3\times 48$ lattices at five different lattice spacings, while at finite temperature we carried out measurements on lattices with $N_t=6,8$ and $10$, to enable a continuum limit extrapolation. We studied finite volume effects on $N_t=6$ lattices, using three different aspect ratios. The light ($m_u=m_d\equiv m_{ud}$) and strange ($m_s$) quark masses are set to their physical values, along the line of constant physics (LCP) as $m_{ud}=m_{ud}(\beta)$, and $m_s/m_{ud}=28.15$.
The LCP was determined by keeping $f_K/M_\pi$ and $f_K/M_K$ physical, and the lattice scale is set using $f_K$.
More details about the lattice action, the determination of the scale and the LCP, and the lattice ensembles can be found in Refs.~\cite{Aoki:2005vt,Borsanyi:2010cj,Bali:2011qj}. At each temperature and external magnetic field we measured the observables of interest on $\mathcal{O}(100)$ thermalized configurations which were separated by 5 trajectories to reduce autocorrelations. The measurements were carried out using the noisy estimator method, with 20--40 random vectors. 

For our choice of the smearing parameters we obtain for the scalar and tensor renormalization constants,
\be
\begin{split}
\ZS^{\MSBar} &= 1 + \frac{g^2C_F}{16\pi^2} \left[ 0.7929 +3 \log\left(a^2\mu^2\right)  \right],\\
\ZT^{\MSBar} &= 1 + \frac{g^2C_F}{16\pi^2} \left[ 1.3136 - \log\left(a^2\mu^2\right)  \right],\\
\end{split}
\label{eq:ZTMSL}
\ee
from Eqs.~(\ref{eq:ZSRI}),~(\ref{eq:ZTRI}) and~(\ref{eq:ZTMS}).
We define the coupling $g$ in the ``E'' scheme~\cite{Bali:1992ru}, using the plaquette expectation value,
\be
g^2_E = \frac{1}{c} \left(1 - \frac{1}{3} \expv{\tr \, U_\square} \right),
\ee
which is found to be $10-20\%$ larger than the bare coupling $g^2$.
We compute $c$ perturbatively for the tree-level improved Symanzik gauge action and obtain $c=0.183131340(2) \cdot C_F$, thereby confirming Ref.~\cite{Bali:2002wf}. 

We allow for a systematic error of $50\%$ in $1-\ZT^{\MSBar}$ for the effect of higher order terms in the perturbative calculation.

\section{Results}
\label{sec:results}

We measure the tensor polarizations as functions of the external field at various temperatures for the three different flavors. We observe that $\expv{\bar\psi_u\sigma_{xy}\psi_u}$ is negative, indicating that $\chiF_u<0$, in accordance with Ref.~\cite{Ball:2002ps} and the discussion about the sign convention below Eq.~(\ref{eq:pbpTdef}). 
Whether this corresponds to a para- or a diamagnetic response will be discussed in Sec.~\ref{sec:magnsusc}.

\begin{figure}[h!]
\includegraphics*[width=9.5cm]{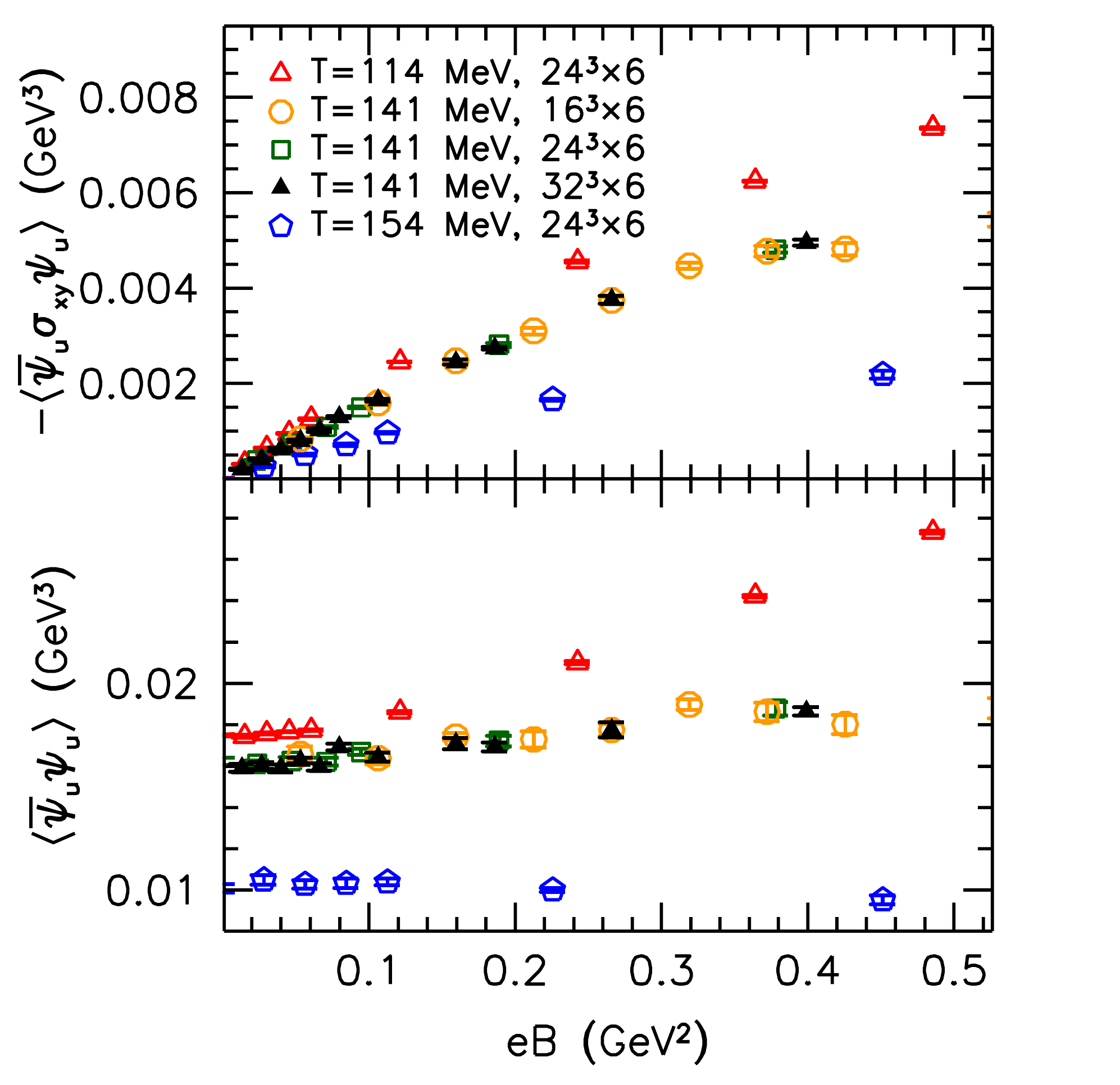}
\vspace*{-0.6cm}
\caption{Minus the bare tensor polarization (upper panel) and the bare condensate (lower panel) for the up quark, for three temperatures on the $N_t=6$ lattices.}
\label{fig:sigmaxy}
\end{figure}

In the upper panel of Fig.~\ref{fig:sigmaxy} we show the bare tensor polarization as a function of the magnetic field for $N_t=6$. We confirm the linear trend to leading order in $B$, in agreement with Ref.~\cite{Buividovich:2009ih}. However, the slope at small $B$ is also observed to change significantly with temperature. We find that nonlinear effects are always below $5\%$ for magnetic fields $eB<0.2 \textmd{ GeV}^2$ and they reduce as the temperature decreases.
In the lower panel of Fig.~\ref{fig:sigmaxy} we also show how the bare condensate itself changes with $B$ for different temperatures. We observe that the dependence of the condensate on $B$ varies strongly with the temperature in the transition region. This behavior was found to be the reason for the decrease of the chiral transition temperature with growing $B$, and was investigated in detail in Refs.~\cite{Bali:2011qj,Bali:2011uf,Bali:2012zg}.
We study finite volume effects at one temperature $T=141$ MeV, for $N_t=6$ ensembles with $N_s=16,24$ and $32$, see Fig.~\ref{fig:sigmaxy}. The largest lattice corresponds to a linear extent of $7$ fm. 
Since we see no deviation for the tensor polarization or the condensate between the different volumes, we conclude that finite size effects are smaller than our statistical errors.

Next, we concentrate on the leading linear trend in $\expv{\bar\psi_f \sigma_{xy} \psi_f}$, i.e. on the slope characterized by the tensor coefficient $\TAU$, as defined in Eq.~(\ref{eq:defchi}). 
We perform the multiplicative renormalization of $\TAU$ according to Eq.~(\ref{eq:mdmsub}), using the tensor renormalization constant, Eq.~(\ref{eq:ZTMSL}) to the $\MSBar$ scheme at a renormalization scale $\mu=2 \textmd{ GeV}$. The dependence of the results on the renormalization scale $\mu$ is found to be mild, as can be seen below.

\begin{figure}[ht!]
\vspace*{-0.2cm}
\includegraphics*[width=9.0cm]{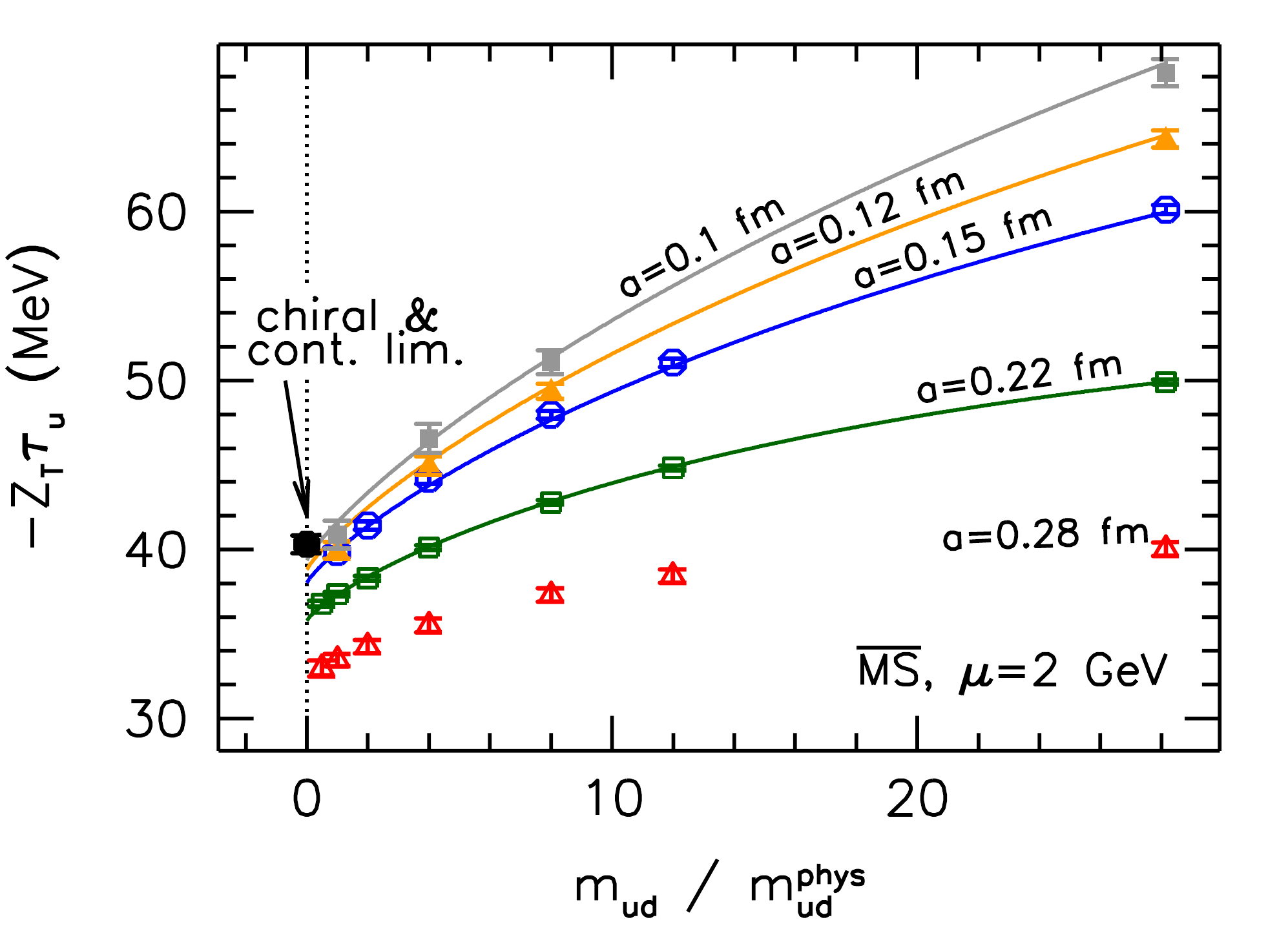}
\vspace*{-0.6cm}
\caption{Mass dependence of the combination $-\ZT\cdot \tau_u$ in the $\MSBar$ scheme at renormalization scale $\mu=2\textmd{ GeV}$. The coefficient of the logarithmic divergence is determined by fitting the data by a lattice spacing-dependent function (solid lines).}
\label{fig:chiral_cont}
\end{figure}

We measure $\ZT\cdot \TAU$ at zero temperature for several lattice spacings and quark masses. Here we fix the strange quark mass to its physical value and tune only the light mass such that $R\equiv m_{ud}/m_{ud}^{\rm phys}$ varies between $0.5$ and $28.15$. For the latter ratio all three quarks have equal masses. (Note that these measurements are also fully dynamical and no partial quenching is applied.) In Fig.~\ref{fig:chiral_cont} we plot minus the tensor coefficient for the up quark as a function of $R$ for five different lattice spacings. Motivated by the behavior of the tensor coefficient in the free case, Eq.~(\ref{eq:polarfree}), and by the scaling properties of the action we use, we consider the following fit function for $\ZT\TAU$:
\be
c_{f0} + c_{f1} R + c_{f2} R \log(R^2a^2),\quad 
 c_{fi}=c_{fi}^{(0)}+c_{fi}^{(1)}a^2.
\label{eq:fitfunc}
\ee
Here $a$ is to be understood in units of $\textmd{GeV}^{-1}$. This form describes the data very well; we obtain $\chi^2/{\rm dof.}\le 1.5$ for both the up and down flavors. The fitted values for $c_{fi}^{(j)}$ are listed in Table~\ref{tab:cs}. We remark that the coefficients of the logarithms, $c_{u2}^{(0)}/m_{ud}^{\rm phys}=0.055(5)$ and $c_{d2}^{(0)}/m_{ud}^{\rm phys}=0.072(6)$ are quite close to the free-field value of $3/(4\pi^2)$ (see Appendix~\ref{app:mlogm}). 
We perform the fit both for all lattice spacings and for only the finest four lattices. Moreover we consider the inclusion of an $R^2$ term in the fit and vary the fit range to exclude points with largest masses. The difference between these fits is used to estimate the systematic error of this combined extrapolation.

\setlength{\tabcolsep}{3pt}
\setlength{\extrarowheight}{3pt}
\begin{table}[ht!]
\begin{tabular}{|c||c|c|c|c|c|c|}
\hline
$f$ & $c_{f0}^{(0)}$ & $c_{f0}^{(1)}$ & $c_{f1}^{(0)}$ & $c_{f1}^{(1)}$ & $c_{f2}^{(0)}$ & $c_{f2}^{(1)}$ \\
\hline
$u$ & -40.3 & 3.8 & -2.1 & 0.5 & 0.19 & -0.03 \\
\hline
$d$ & -38.9 & 2.8 & -2.5 & 0.7 & 0.25 & -0.07 \\
\hline
\end{tabular}
\caption{Central values for the fit parameters of Eq.~(\protect \ref{eq:fitfunc}) in units of $\textmd{MeV}$.}
\label{tab:cs}
\end{table}

At zero quark mass the additive divergence is absent and therefore, applying the combined fit, the continuum limit of the chiral limit of $\ZT\TAU$ can be extracted (it equals the $c_{f0}^{(0)}$ parameter). This corresponds to the black point in Fig.~\ref{fig:chiral_cont}. 
However, since we are interested in the tensor coefficient at physical quark masses, we now follow the scheme of Eq.~(\ref{eq:mdmsub}), subtracting the logarithmic divergence. We apply the operator $1-m_f\partial_{m_f}=1-R\partial_R$, which acting on the fit function of Eq.~(\ref{eq:fitfunc}) yields 
\be
\TAU^r=c_{f0} -2 c_{f2} R.
\label{eq:fitfunc_actedon}
\ee
As already emphasized in Sec.~\ref{sec:renormalization}, the subtraction of the divergent term $\TAU^{\rm div}$ does not affect the chiral continuum limit since it vanishes at $m_f=0$. Moreover, this subtraction eliminates the scheme-dependent finite terms (cf. Eq.~(\ref{eq:polarfreefinal})), making the conversion to the $\MSBar$ scheme trivial.

For the strange quark we do not perform a similar analysis with modified strange quark masses, but subtract the logarithmic divergence by using the fit parameters for the down quark and $R=28.15$. We find that the dependence of the strange quark tensor polarization on the light quark masses is below a few per cent ($1\%$ for the coarsest and $4\%$ for the finest lattice). Therefore this approximation introduces errors smaller than those already present due to statistics and renormalization.

After the subtraction, the renormalized tensor coefficient $\TAU^r$ has a well defined continuum limit even for finite quark masses. 
We find that for physical light quark masses $|\tau_{u,d}^{\rm div}|<2.5 \textmd{ MeV}$ for our range of lattice spacings. For the strange quark the divergent contribution is larger in magnitude, giving rise to larger errors due to this subtraction.

Our final results for the zero temperature renormalized tensor coefficients in the $\MSBar$ scheme at a renormalization scale $\mu=2 \textmd{ GeV}$ are summarized in Table~\ref{tab:zeroTres}. For the light flavors this Table contains the results both for physical quark masses and for the chiral limit.
These values may be compared to the unrenormalized quenched $\mathrm{SU}(2)$ lattice result $-\tau_{ud}=46(3)$ MeV of Ref.~\cite{Buividovich:2009ih} 
and to a similar study in the quenched $\mathrm{SU}(3)$ theory, $52 \textmd{ MeV}$~\cite{Braguta:2010ej}. Our results are in reasonable agreement with the QCD sum rule result $50(15)$ MeV of Ref.~\cite{Ball:2002ps}, which was calculated at $\mu=1 \textmd{ GeV}$ (note that the scale dependence of $\tau_f$ is small due to its small anomalous dimension, see Eq.~(\ref{eq:ZTMSL})). We also compare our results to the NJL and quark-meson model predictions of $69$ MeV and $65$ MeV~\cite{Frasca:2011zn}, respectively, which were obtained at an even lower renormalization scale of $\mu\sim0.6 \textmd{ GeV}$. We remark that the same authors obtain a lower value of $44$ MeV in the renormalized version of the quark meson model~\cite{Frasca:2011zn}.

\setlength{\tabcolsep}{3pt}
\setlength{\extrarowheight}{1pt}
\begin{table}[ht!]
\begin{tabular}{|c|c||c|c|c|c|c|c|}
\hline
\multirow{2}{*}{$f$} & \multirow{2}{*}{$m$} & \multirow{2}{*}{$\tau_f^r$} & \multicolumn{5}{c|}{error} \\ 
\cline{4-8}
& & & stat. & mult. & cont. & scale & total\\
\hline
\multirow{2}{*}{$u$} & phys. & -40.7 & 0.2 & 0.3 & 1.0 & 0.8 & 1.3 \\
\cline{2-8}
& chir. & -40.3 & 0.2 & 0.3 & 1.1 & 0.8 & 1.4 \\
\hline
\multirow{2}{*}{$d$} & phys. & -39.4 & 0.3 & 0.3 & 1.1 & 0.8 & 1.4 \\
\cline{2-8}
& chir. & -38.9 & 0.3 & 0.3 & 1.3 & 0.8 & 1.5 \\
\hline
$s$ & phys. & -53.0 & 0.5 & 0.8 & 7.1 & 1.1 & 7.2 \\
\hline
\end{tabular}
\caption{Results and error budget for the renormalized tensor coefficients for physical quark masses (phys.) and in the chiral limit (chir.). Given are (in units of $\textmd{MeV}$) the errors related to statistics, the multiplicative renormalization, the combined continuum fit, the lattice scale and, finally, the total error.}
\label{tab:zeroTres}
\end{table}

\begin{figure}[ht!]
\vspace*{-0.1cm}
\includegraphics*[width=9.0cm]{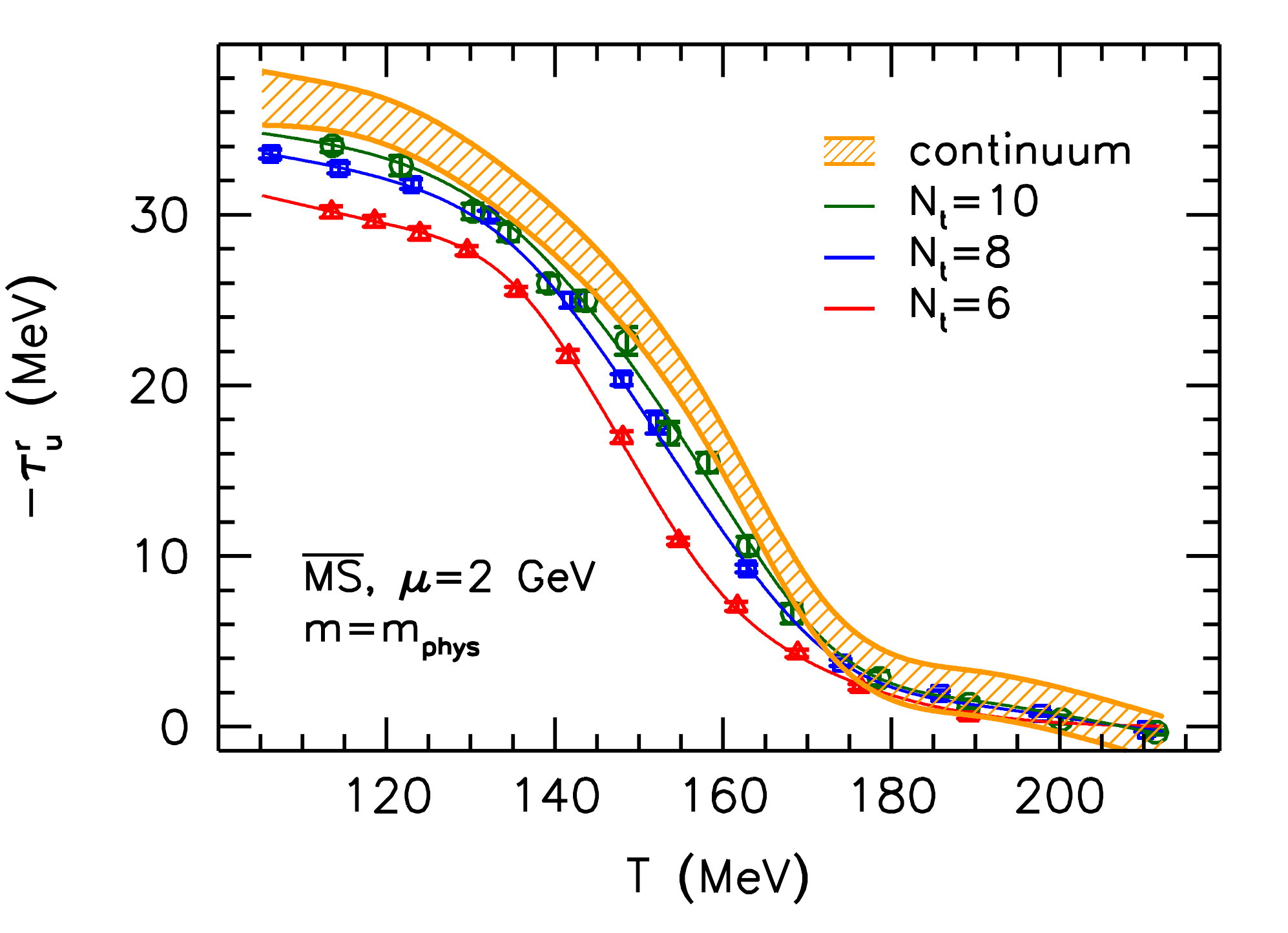}
\vspace*{-0.7cm}
\caption{Minus the renormalized tensor coefficient $\tau_u^r(T)$ in the $\MSBar$ scheme at a renormalization scale $\mu=2 \textmd{ GeV}$ for three lattice spacings and the continuum extrapolation.}
\label{fig:finiteT}
\end{figure}

Next, we use the fact that the $\TAU^{\rm div}$ contribution is independent of $T$ to perform the additive renormalization of the tensor coefficient at finite temperatures. 
In Fig.~\ref{fig:finiteT} we plot $-\tau_u^r$ as a function of the temperature for three lattice spacings. We perform a simultaneous fit of the results for different lattice spacings to an $N_t$-dependent spline function. This dependence is of the form $N_t^{-2}$, once again to reflect the scaling properties of our lattice action. We can read off the continuum extrapolation at $N_t^{-2}=0$, which is shown in the figure by the hatched yellow band. 
The systematic error of the continuum extrapolation is estimated to be $1 \textmd{ MeV}$ based on our experience at $T=0$ (see Table~\ref{tab:zeroTres}) and is added to the statistical error in quadrature. Moreover, the uncertainty in the determination of the lattice scale (for details see Ref.~\cite{Borsanyi:2010cj}) propagates into this result and gives rise to an additional systematic error of $2\%$. Since this latter error is uniform and does not influence the shape of the $\tau^r_f(T)$ curve, it is not included in the plot.

In the same manner we determine the tensor coefficient for the down quark at $T>0$, and obtain results which are within errors consistent with $\tau^r_u$, just as was observed at $T=0$. For the strange quark this procedure leads to a qualitatively similar temperature-dependence too.
The dependence of $\tau_{u,d}^r$ on the temperature in the transition region can be used to define a transition temperature at $B=0$. We determine the inflection point of $\tau_{u,d}^r(T)$ and obtain $T_c=162(3)(3) \textmd{ MeV}$ in the continuum limit. Here the first error combines the statistical error and the error coming from the continuum extrapolation, and the second one is due to the uncertainty in the lattice scale. In conclusion, the tensor coefficient acts as a quasi-order parameter for the chiral transition, and gives a similar transition temperature as the chiral condensate at $B=0$, $T_c=159(3)(3) \textmd{ MeV}$, cf. Refs.~\cite{Bali:2011qj,Borsanyi:2010bp}.

Finally, we study the dependence of $\tau_f^r$ on the renormalization scale $\mu$ at $T=0$. We carry out the analysis for a range of renormalization scales in the window $1 \textmd{ GeV}\le \mu \le 4\textmd{ GeV}$. We find a very mild dependence on $\mu$ such that the tensor coefficients remain within the total errors given in Table~\ref{tab:zeroTres}.

\section{Magnetic susceptibility}
\label{sec:magnsusc}

We can translate the result for $\TAU^r$ to the magnetic susceptibility $\chiF_f$ of Eq.~(\ref{eq:defchi}) using the (scale- and scheme-dependent) value of the quark condensate. 
We recall the Gell-Mann-Oakes-Renner relation,
\be
M_\pi^2 F^2 = (m_u + m_d) \cdot \expv{\bar\psi_l\psi_l}+\ldots,
\ee
which, at zero external field and in the chiral limit, relates the light condensate $l=u,d$ to the quark masses and to the pion mass and decay constant, with $F=86.2 (5)$ MeV~\cite{Colangelo:2003hf}. We make use of a recent lattice determination~\cite{Durr:2010vn} of the quark masses in the $\MSBar$ scheme at $\mu=2\textmd{ GeV}$, 
$m_u+m_d=6.94(13) \textmd{ MeV}$, to extract $\expv{\bar\psi_l\psi_l} = (269 (2)\textmd{ MeV})^3$. 
(We mention that multiplying the lattice bare mass along the LCP~\cite{Borsanyi:2010cj} and the inverse of the scalar renormalization constant of Eq.~(\ref{eq:ZTMSL}), we get a compatible value for the renormalized quark mass in the $\MSBar$ scheme, albeit with large uncertainties.) 
For the strange condensate we employ the QCD sum rule prediction~\cite{Jamin:2002ev}, $\expv{\bar\psi_s\psi_s}/\expv{\bar\psi_l\psi_l}=0.8(3)$.
Using these values for the quark condensates, the zero-temperature magnetic susceptibilities at physical quark masses are calculated as
\be
\MSBar,\; \mu=2 \textmd{ GeV}:\hspace*{-2.8cm}
\begin{split}
\chiF_u &= - (2.08\pm0.08) \textmd{ GeV}^{-2}, \\
\chiF_d &= - (2.02\pm0.09) \textmd{ GeV}^{-2}, \\
\chiF_s &= - (3.4\pm1.4) \textmd{ GeV}^{-2}.
\end{split}
\hspace*{-2cm}
\label{eq:chiresult}
\ee
The magnetic susceptibilities at different values of $\mu$ can be obtained by running down with the ratio of renormalization constants $\ZT^{\MSBar}/\ZS^{\MSBar}$. Using the four-loop running to $\mu=1\textmd{ GeV}$ one has to multiply the above values by $r=1.49(7)$. 
We remark furthermore that running down with $\ZS^{\MSBar}$ to a renormalization scale of $\mu=1 \textmd{ GeV}$ we obtain $\expv{\bar\psi_l\psi_l} = (245 (5)\textmd{ MeV})^3$.

Our results in Eq.~(\ref{eq:chiresult}) are in good agreement with the QCD sum rule calculations\footnote{The value given in Ref.~\cite{Ball:2002ps} is $\chiF_l=3.15(30) \textmd{ GeV}^{-2}$ at $\mu=1 \textmd{ GeV}$. We divided this by $-1.49(7)$, running the value to the scale $\mu = 2 \textmd{ GeV}$ and accounting for the different sign convention we employ, see the remark after Eq.~(\ref{eq:pbpTdef}).}
 summarized and updated in Ref.~\cite{Ball:2002ps}: $\chiF_l = -2.11(23) \textmd{ GeV}^{-2}$ at $\mu= 2 \textmd{ GeV}$, and also compare well with the vector dominance estimate of $\chiF_l=-2/m_\rho^2\approx -3.3\textmd{ GeV}^{-2}$. 
We remark that for the strange susceptibility, QCD sum rules predict $\chiF_s\approx \chiF_l$~\cite{Balitsky:1997wi}, which is somewhat smaller than our result in Eq.~(\ref{eq:chiresult}).

Comparing the temperature-dependence of the light tensor coefficient (Fig.~\ref{fig:finiteT}) and that of the light quark condensate from Ref.~\cite{Bali:2012zg}, we conclude that the ratio of the two renormalized observables is compatible with a constant, resulting in a magnetic susceptibility $\chiF_l(T)$ depending only weakly on the temperature, at least for temperatures $T<170\textmd{ MeV}$.
Moreover, we remark that since $\chiF_f$ is given in terms of the chiral condensate (which has a large anomalous dimension), the magnetic susceptibility has a stronger scale dependence than $\TAU^r$. 

As anticipated in the introduction, the magnetic susceptibility $\chiF_f$ of the condensate is intimately connected to the spin contribution $\chiM^S$ to the total magnetic susceptibility. Using this equivalence (which we prove in Appendix~\ref{app:logZB}), one sees that with our sign conventions $\chiF_f>0$ corresponds to paramagnetism and $\chiF_f<0$ to diamagnetism. Thus we conclude that the response of the QCD quark condensate to external magnetic fields is in its nature {\it diamagnetic}.

\section{Conclusions}
\label{sec:conclusions}

In this paper we studied the response of the QCD vacuum to a constant external magnetic field at zero and at finite temperature. We determined the tensor polarizations of the quark condensates for various temperatures and external fields. We observed that the polarization of the flavor $f$ at a temperature $T$ is a linear function of $B$ for fields $eB<0.2 \textmd{ GeV}^2$, with a coefficient $\TAU(T)$, defined in Eq.~(\ref{eq:defchi}). The renormalization of this tensor coefficient requires two steps. The additive divergences (which are present for finite quark masses) were fitted explicitly at $T=0$ and then subtracted using the operator $1-m\partial_m$, at $T=0$ and at $T>0$. The multiplicative renormalization was performed perturbatively. We obtained results in the $\MSBar$ scheme at a renormalization scale $\mu= 2 \textmd{ GeV}$, and extrapolated these to the continuum limit using several lattice spacings. Our final results for the renormalized $\TAU^r$ are given in Table~\ref{tab:zeroTres} for $T=0$ and are shown in Fig.~\ref{fig:finiteT} for $T>0$. 
Combining the results for $\TAU^r$ and the quark condensates we also determined the magnetic susceptibilities $\chiF_f$, see Eq.~(\ref{eq:chiresult}) for the zero temperature values. We found $\chiF_f$ to remain constant within errors as the temperature is increased up to $T\approx170 \textmd{ MeV}$.

We showed furthermore that there is a simple relation between the tensor coefficients $\TAU^r$ and the spin contribution $\chiM^S$ to the total magnetic susceptibility, see Eq.~(\ref{eq:therelation}). The negative sign of $\chiM^S$ reveals a diamagnetic response, i.e., that the spin magnetization of the medium aligns itself antiparallel to the external field. 
The magnitude of this effect reduces as the temperature grows, as $\chiM^S$ is proportional to $\TAU^r$, plotted in Fig.~\ref{fig:finiteT}.
For the free case $\chiM^S$ and $\chiM^L$ are known to have opposite signs~\cite{Nielsen:1980sx}, implying a partial cancellation between the two sectors. Therefore, a determination of the
orbital angular momentum contribution is necessary to arrive at a definite conclusion whether
the total response of the QCD vacuum to external magnetic fields is para- or diamagnetic. 

\vspace*{0.4cm}
\noindent
{\bf Acknowledgments. } This work has been supported by the EU FP7 grants ERC no. 208740 and PITN-GA-2009-238353 (ITN STRONGnet) and by the DFG (BR 2872/4-2 and SFB/TR 55).
Computations were carried out on the GPU clusters~\cite{Egri:2006zm} at the E\"otv\"os University Budapest and at the University of Regensburg.
M.~Constantinou acknowledges financial support from the Cyprus Research Promotion Foundation under contract number TECHNOLOGY/$\Theta$E$\Pi$I$\Sigma$/0308(BE)/17. The authors would like to thank Vladimir Braun and Meinulf G\"ockeler for essential remarks and Zolt\'an Fodor for a careful reading of the manuscript. G.~Endr\H{o}di would like to thank Jens Andersen, Pavel Buividovich, Tommy Burch and Marco Ruggieri for useful discussions. 

\appendix

\section{Spin- and orbital angular momentum- contributions}
\label{app:logZB}

The partition function of QCD (this time without taking the root of the determinant as in the staggered lattice formulation) is given by the functional integral,
\be
\ZZ = \int \D U e^{-\beta S_g} \prod_f \det (\dsf+m_f),
\label{eq:appZ}
\ee
with the massless Dirac operator $\dsf=\gamma_\mu D_{\mu,f}$ and covariant derivative $D_{\mu,f}=\partial_\mu+iq_fA_\mu + iA_\mu^a T^a$. For an external magnetic field in the $z$-direction one has $\partial_x A_y-\partial_y A_x=B$ and $A_z=A_t=0$. 

The derivative of the logarithm of Eq.~(\ref{eq:appZ}) with respect to $B$ is
\be
\frac{\partial \log\ZZ}{\partial B} = \sum_f \expv{\tr \frac{1}{\dsf+m_f} \frac{\partial \dsf}{\partial B}}.
\ee
We manipulate this using $\tr\, \partial \dsf/\partial B \propto\tr\, \gamma_\mu=0$ and the cyclicity of the trace:
\be
\begin{split}
\frac{\partial \log \ZZ}{\partial B}&=\sum_f\frac{1}{m_f} \expv{ \tr \left(\frac{m_f}{\dsf+m_f}-1\right)\frac{\partial \dsf}{\partial B}}\\
 &=-\sum_f\frac{1}{m_f}\expv{ \tr \frac{1}{\dsf+m_f}\dsf\frac{\partial \dsf}{\partial B}}\\
 &=-\frac{1}{2}\sum_f\frac1{m_f}\expv{ \tr \frac{1}{\dsf+m_f}\frac{\partial \dsf^2}{\partial B}}.
\end{split}
\ee
The derivative of the square of the Dirac operator in the magnetic field background, after a standard simplification involving $\gamma$-matrices, reads
\be
\frac{\partial \dsf^2}{\partial B} = \frac{\partial D_{f}^{2}}{\partial B}- q_f \sigma_{xy},
\label{eq:D2der}
\ee
where $D_f^2 = D_{\mu,f}D_{\mu,f}$ with summation for $\mu$ but not for $f$. 
This implies,
\be
\frac{T}{V}\frac{\partial \log\ZZ}{\partial B} =\frac{1}{2}\sum_f\frac{q_f}{m_f} \left(\expv{\bar\psi_f \sigma_{xy}\psi_f} + \expv{\bar\psi_f L_{xy}\psi_f} \right),
\ee
where we defined
\be
L_{xy}\equiv-\frac{\partial D_{f}^{2}}{\partial(q_f B)}.
\label{eq:Lxydef}
\ee
This operator corresponds to a generalized angular momentum, as for the choice $A_x=-By/2,\,A_y=Bx/2$ (such that $\partial_\mu A_\mu=0$), it assumes the form $L_{xy}= -i(x\partial_y-y\partial_x)+q_fB(x^2+y^2)/2-yA_x^aT^a+xA_y^aT^a$.

Altogether, using the definition of the (total) magnetic susceptibility, Eqs.~(\ref{eq:defchitotal}) and~(\ref{eq:chiSO}), we get
\be
\chiM_f = \left.\frac{q_f/e}{2m_f}  \left( \frac{\partial \expv{\bar\psi_f\sigma_{xy}\psi_f}}{\partial (eB)} + \frac{\partial \expv{\bar\psi_f L_{xy} \psi_f} }{\partial (eB)} \right)\right|_{eB=0},
\label{eq:xicontr}
\ee
showing two separate contributions $\chiM^S+\chiM^L$ to the total susceptibility, cf.\ Eq.~(\ref{eq:therelation}). 

The conventional calculation of the spin- and orbital momentum-related contributions to $\chiM$ yields the same result. Below we demonstrate this for the free case.
Here the spin-related contribution to the change in the free energy density due to the magnetic field at zero temperature is given by~\cite{Nielsen:1980sx,Wilczek:1996bw,Grozin:2008yd},
\bea
\Delta f^S &&= -N_c \int  \frac{ d^3 p}{(2\pi)^3}\! \\
 && \times \sum_{f,s=\pm1} \left( \sqrt{p^2 + m_f^2 + s\,q_fB} -\sqrt{p^2+m_f^2}\right). \nonumber
\eea
Employing the definition of the total susceptibility, Eq.~(\ref{eq:defchitotal}), the spin-dependent contribution equals
\be
\chiM^S \!=\! -\!\left.\frac{\partial^2 \Delta f^S}{\partial (eB)^2}\right|_{eB=0}
\!\!=\! -N_c\sum_f \frac{(q_f/e)^2}{2\pi} \!\int\!\! \frac{d^2p}{(2\pi)^2} \frac{1}{p^2+m_f^2}.
\label{eq:xiS}
\ee
In Appendix~\ref{app:mlogm} we will calculate the tensor polarization in the free case.
Comparing Eq.~(\ref{eq:xiS}) with Eq.~(\ref{eq:polarfreepre}) below, we see that the first term of Eq.~(\ref{eq:xicontr}) is indeed the spin-related contribution, $\chiM^S$. The second term of Eq.~(\ref{eq:xicontr}) is then identified with the orbital momentum coupling.
The two contributions to Eq.~(\ref{eq:chiSO}) then read,
\be
\chiM^S = \sum_f\frac{(q_f/e)^2}{2m_f} \TAU, \quad\quad
\chiM^L = \sum_f\frac{q_f/e}{2m_f} \frac{\partial \expv{\bar\psi_fL_{xy}\psi_f}}{\partial (eB)},
\label{eq:chiSandtau}
\ee
where we used the definition of the tensor coefficient, Eq.~(\ref{eq:defchi}).
This shows that the tensor coefficient of the quark condensate is responsible for the spin contribution of the total magnetic susceptibility. 
Recalling the relation between the sign of $\chiM^S$ and para/diamagnetism as discussed in the introduction, we conclude that with our sign conventions $\tau_f>0$ ($\chiF_f>0$) corresponds to paramagnetism, while $\tau_f<0$ ($\chiF_f<0$) to diamagnetism.
We remark that on the lattice $\chiM^L$ cannot directly be computed from Eq.~(\ref{eq:Lxydef}), due to the quantization of the magnetic flux.

\section{Logarithmic divergence in the tensor polarization}
\label{app:mlogm}

In this appendix we will demonstrate the appearance of a logarithmic divergence in the tensor polarization of the condensate. We consider one free quark with electric charge $q_f$ and mass $m_f$ at vanishing temperature.

The negative square of the Dirac operator in the  background of a constant magnetic field is well-known to have eigenvalues~\cite{springerlink:10.1007/BF01397213,Heisenberg:1935qt}
\be
\begin{split}
 -\dsf^2 \to \lambda^2&= p_0^2+p_z^2+(2n+1)|q_fB|+s\,q_fB,
\end{split}
\ee
being twice degenerate (incorporating particle and antiparticle). Here $p_0,\,p_z$ are momenta, $n=0,\,1,\,\ldots$ labels the Landau levels and $s=\pm 1$ is twice the spin (these are the eigenvalues of $\sigma_{xy}$), which is coupled to the magnetic field (here we do not consider anomalous magnetic moments). The sum over the eigenvalues is performed according to (see e.g. Ref.~\cite{Nielsen:1980sx}),
\be
\sum_{\lambda^2}=2N_c\,\frac{1}{2\pi T}\int_{-\infty}^\infty \!\!\!dp_0\,
 \frac{L_z}{2\pi}\int_{-\infty}^\infty \!\!\!dp_z\,
 \frac{L_xL_y |q_fB|}{2\pi}\sum_{n=0}^\infty\sum_{s=\pm 1}.
\ee

For the tensor polarization of Eq.~(\ref{eq:pbpTdef}) we note that due to chirality (and since $\gamma_5$ commutes with $\sigma_{xy}$),
\bea
 \tr \frac{1}{\dsf+m_f}\sigma_{xy}&=&\tr \,\gamma_5\frac{1}{\dsf+m_f}\gamma_5\sigma_{xy}\\
  &=&\tr \frac{1}{-\dsf+m_f}\sigma_{xy}
  =m_f\,\tr \frac{\sigma_{xy}}{-\dsf^2+m_f^2},\notag
\eea
which results in the spectral representation (omitting the staggered factor of $1/4$),
\bea
&&\expv{ \bar\psi_f \sigma_{xy} \psi_f} 
 = N_c\frac{m_f|q_fB|}{\pi}\!\int\! \frac{d^2p}{(2\pi)^2}\\ 
 &&\quad\quad\times  \sum_{n,s}\frac{s}{p^2+\big(2n+1+s\,\sign(q_fB)\big)|q_fB|+m_f^2}.\notag
\eea
In the sum the contributions $\{n=k,s\,\sign(q_fB)=1\}$ and $\{n=k+1,s\,\sign(q_fB)=-1\}$ cancel leaving only the unpaired lowest Landau level $\{n=0,s\,\sign(q_fB)=-1\}$, as was also noted in Ref.~\cite{Frasca:2011zn}. Hence we get
\be
\expv{ \bar\psi_f \sigma_{xy} \psi_f} 
 = -N_c\frac{m_f\,q_fB}{\pi}\int\!\frac{d^2p}{(2\pi)^2}\frac{1}{p^2+m_f^2}.
\label{eq:polarfreepre}
\ee
This cancellation can be confirmed via zeta function regularization and is absent for other observables like the free energy or the condensate. As the eigenvalue of the lowest Landau level is $B$-independent, the free tensor polarization is exactly linear in the magnetic field.

We evaluate the remaining logarithmically divergent integral with dimensional regularization in $d=2-\epsilon$ dimensions,
\be
\expv{ \bar\psi_f \sigma_{xy} \psi_f} 
\! =\! N_c\frac{m_f\,q_fB}{4\pi^2}\!\left[-\frac{2}{\epsilon}+\gamma+\log\left(\frac{m_f^2}{4\pi}\right)\right] +\mathcal{O}(\epsilon).
\label{eq:polarfree}
\ee
A $\log m_f^2$-term has appeared, whose coefficient is scheme-independent, for 3 colors it is $3/(4\pi^2)\cdot m_f\,q_fB$ (cf. Ref.~\cite{Ioffe:1983ju} with different sign conventions). Also the singularity for $\epsilon\to 0$ has been isolated and can be subtracted through a particular renormalization scheme, introducing a cut-off $\Lambda$ such that $\expv{ \bar\psi_f \sigma_{xy} \psi_f}\propto \log (m_f^2/\Lambda^2)$, or, on the lattice $\log (m_f^2a^2)$. The finite term ($\gamma-\log(4\pi)$ in Eq.~(\ref{eq:polarfree})) is scheme-dependent (in our lattice scheme it reads $0.1549\,\pi^2 -\log4$) but, together with the logarithmic contribution, it disappears from the combination
\be
(1-m_f\partial_{m_f})\expv{ \bar\psi_f \sigma_{xy} \psi_f} 
 = -\frac{m_f\,q_fB}{2\pi^2},
\label{eq:polarfreefinal}
\ee
as we also emphasized in the body of the paper, Eq.~(\ref{eq:fitfunc_actedon}). Note that $(1-m_f\partial_{m_f})$ acting on Eq.~(\ref{eq:polarfreepre}) renders the integral finite and allows for a direct computation of the coefficient of the logarithmic term.

\section{Definition of staggered operators}
\label{app:staggeredops}

To discretize the continuum tensor bilinear operator on the lattice in the staggered formalism, 
one has to define fields that live on corners of four-dimensional elementary hypercubes of the lattice. The position 
of a hypercube inside the lattice is denoted by the index $y$, where $y$ is a
four-vector with even components $y_\mu=0,2,\ldots, N_\mu-2$ (all lattice extents are even). 
The position of a fermion field component within a specific hypercube is defined by one additional 
four-vector index, $C$ ($C_\mu \in \{0,1\}$). 
Following the notation of Ref.~\cite{Patel:1992vu}, this hypercube field is denoted in this section by $\chi(y)_C=\chi(y+C)/4$ 
(instead of $\psi(x)$, to distinguish between the different ways of labeling).
After rotating into the staggered basis, the operator ${\cal O}_s=\bar\psi_f\Gamma_s \psi_f$
can be written as~\cite{Patel:1992vu}
\be
{\cal O}_s = \sum_{C,D}
\bar\chi(y)_C\,\left(\overline{\Gamma_s	\otimes\openone}\right)_{CD}\,\chi(y)_D,
\label{eq:tensorop}
\ee
where the matrix $\Gamma_s$ acts on spin and the unit matrix $\openone$ on taste components of the
staggered field $\chi(y)$ (we will consider the scalar $\Gamma_S=\openone$ and the tensor $\Gamma_T=\smn$ spin structures). 
Here 
\be
\left(\overline{\Gamma_s\otimes\openone}\right)_{CD} \equiv 
\frac{1}{4}\,{\rm Tr}\left[\gamma^\dagger_C\,\Gamma_s\,\gamma_D\,\openone\right],
\label{eq:gamma1}
\ee
with the staggered transformation given by
\be
\gamma_C\equiv\gamma_1^{C_1}\,\gamma_2^{C_2}\,\gamma_3^{C_3}\,\gamma_4^{C_4}.
\ee

The operator of Eq.~(\ref{eq:tensorop}) is clearly not gauge invariant, since $\bar\chi$ and $\chi$ are defined in
different points of the hypercube. To restore gauge
invariance, we insert the average of products of gauge link variables
along all possible shortest paths connecting the sites $y+C$ and
$y+D$~\cite{Patel:1992vu}. This average is denoted by $U_{C,D}$ and the gauge invariant
operator is now
\be
{\cal O}_s = \sum_{C,D} \bar\chi(y)_C\,
\left(\overline{\Gamma_s\otimes\openone}\right)_{CD}\,
U_{C,D}\,\chi(y)_D\,.
\label{eq:Oper}
\ee
Using the definition of Eq.~(\ref{eq:gamma1})
we can further simplify the expression for the operator ${\cal O}_s$.
One useful identity is
\be
\gamma_{\mu}\,\gamma_{C} = \gamma_{C+\hat{\mu}}\,\eta_{\mu}(C)\,,\qquad\eta_{\mu}(C) \equiv (-1)^{\sum_{\nu<\mu}\!C_\nu},
\ee
so that
\be
\begin{split}
\frac{1}{4}{\rm Tr}\left[\gamma_C^\dagger\,\openone\,\gamma_D\right]
&= \delta_{C,D}\,,\\
\frac{1}{4}{\rm Tr}\left[\gamma_C^\dagger\,\smn\,\gamma_D\right]
&=\frac{1}{i}\, \delta_{C,D+\hat{\mu}+\hat{\nu}}\,\eta_\nu(D)\,\eta_\mu(D+\hat{\nu}).
\end{split}
\label{eq:Oper2}
\ee
(Here and below, in expressions such as $C+\hat{\mu}$ the sum is to be taken modulo 2.)
Using Eq.~(\ref{eq:Oper2}), the scalar and tensor operator (the latter for any $\mu\neq\nu$) can be written as
\be
{\cal O}_S = \sum_D
\bar\chi(y)_{D} \,
\chi(y)_D\, ,
\label{eq:OS2}
\ee
\be
\begin{split}
{\cal O}_T  = \frac{1}{i} \sum_D\,
&\bar\chi(y)_{D+\hat{\mu}+\hat{\nu}}\,U_{D+\hat{\mu}+\hat{\nu},D} \, \chi(y)_D \\
&\times \eta_\nu(D) \,\eta_{\mu}(D+\hat{\nu}).
\end{split}
\label{eq:OT2}
\ee
As can be seen from the above equation, the tensor operator contains a distance of two links (in orthogonal directions) between the
fermion and the antifermion fields. Thus, the product of gauge links entering Eq.~(\ref{eq:OT2}) is
\be
U_{D+\hat{\mu}+\hat{\nu}, D} = \frac{1}{2}\left[\,
U_{D+\hat{\mu}+\hat{\nu}, D+\hat{\mu} } \,U_{D+\hat{\mu}, D}
+ \{ \mu\leftrightarrow \nu\}
\,\right].
\label{eq:paths}
\ee
For the implementation of the tensor Dirac structure we use a modified version of the corresponding part of the MILC code~\cite{milc7.6}.
\vfill

\bibliographystyle{jhep}
\bibliography{magnsusc}

\end{document}